\documentclass{aa}

\usepackage{natbib}
\usepackage{epsfig}

\begin{document}

\title{Population synthesis as a probe of neutron star thermal evolution}

\titlerunning{Population synthesis and neutron star thermal evolution}

\author{S. Popov
         \inst{1}
          \and
H. Grigorian
            \inst{2,3}
       \and
R. Turolla
           \inst{4}
\and
D. Blaschke
            \inst{5,6}
\thanks{Present address: GSI, D-64291 Darmstadt, Germany}
         }
\authorrunning{S. Popov et al.}
   \offprints{D. Blaschke}

\institute{
Sternberg Astronomical Institute,
Universitetski pr. 13, 119992 Moscow, Russia
%\email{polar@sai.msu.ru}
              \and
Department of Physics, Yerevan State University, Alex
        Manoogian Str. 1, 375025 Yerevan, Armenia
\and
Institut f\"ur Physik, Universit\"at Rostock,
D-18051 Rostock, Germany
      \and
Universit\`a di Padova, Dipartimento di Fisica,
via Marzolo 8, 35131 Padova, Italy
            \and
Fakult\"at f\"ur Physik, Universit\"at Bielefeld,
D-33615 Bielefeld, Germany\\
\email{blaschke@theory.gsi.de}
\and
Bogoliubov Laboratory of Theoretical Physics,
Joint Institute for Nuclear Research, 141980 Dubna, Russia}

\date{}

\abstract{The study of thermal emission from isolated, cooling neutron
stars plays a key role in probing the physical conditions of both the
star crust and the core.
The comparison of theoretical models for the star thermal evolution with
the surface temperature derived from X-ray observations of sources of
different age is one of the main tools to investigate the properties
of the interior and constrain the equation of state. Here we propose to use
population synthesis studies as an independent approach to test the physics
governing the star cooling. Theoretical $\mathrm{Log\, N}$-$\mathrm{Log\, S}$
distributions depend on the assumed neutron star thermal evolution. We have
computed distributions for several different cooling scenarios and found that
comparison with the observed $\mathrm{Log\, N}$-$\mathrm{Log\, S}$ of
isolated neutron stars is effective in discriminating among cooling models.
Among the eleven cooling models considered in this paper, all of which
may reproduce the observed temperature vs. age diagram, only
at most three can explain the $\mathrm{Log\, N}$-$\mathrm{Log\, S}$
distribution of close-by cooling neutron stars.
The $\mathrm{Log\, N}$-$\mathrm{Log\, S}$ test, being a ``global'' one and
despite some limitations, appears indeed capable of complementing the
standard temperature vs. age test used up to now.
}

\maketitle

\section{Introduction}
\label{intro}

The determination of the equation of state (EOS) at or above nuclear
densities is a long sought goal in high energy astrophysics. In this
respect direct observations of neutron stars (NSs) may
provide invaluable insight into several key issues of fundamental
physics, like quantum chromodynamics/electrodynamics, superfluidity and
superconductivity, that could not be otherwise tested under laboratory
conditions. An ideal way to place tight constraints on the EOS is
the simultaneous mass and radius measurement of NSs  (see
e.g. \citealt{lapra2001}). Precise mass determinations have been
obtained for NSs in binaries, especially in radio pulsar systems
(e.g. \citealt{tho99}). Simultaneous mass and radius measurements are
presently available for a few X-ray binaries, although masses
derived from gravitational redshift of spectral lines are still
uncertain in many cases (e.g. \citealt{c2002}). Mass estimates can be
obtained also from quasi-periodic oscillation
measurements (\citealt{bul2000}), but results based on this, as well as other
approaches, are still quite model dependent. The study of glitches
observed in radio pulsars, and recently
in anomalous X-ray pulsars (AXPs) as well
(e.g. \citealt{hor2004}; \citealt{osso2003}),
promise further insight
into the understanding of the internal structure of NSs, as well as
future observations of neutrino and gravitational wave emission
from neutron star sources.

For the time being, however, X-ray observations of NSs are,
and will still be for some time,
central for our understanding of the star interior. Isolated
neutron stars that emit at X-ray energies as they
cool are particularly promising in this respect. Their thermal radiation 
directly comes from the star surface, carrying information on the
physical conditions of the emitting matter, in particular on the star
surface temperature. Thermal X-ray emission has been detected from about 20
isolated NSs so far, including normal radio pulsars, central
compact objects in supernova remnants (CCOs in SNRs),
radioquiet NSs, AXPs and soft $\gamma$-repeaters
(see e.g. \citealt{paza03}; \citealt{kaspi04}; \citealt{hab04} for reviews).

Significant progress in the understanding of NS thermal evolution has
been made in recent years and cooling curves have been computed
by several groups (see e.g. \citealt{kam};
\citealt{tsuruta}; \citealt{page}; \citealt{bgv2004} and references therein).
%%%%%%%%%%%%%%%%%%%%%%%%%%%%%%%%%%%%%% Hovik + David insertion
The present work is based on NS cooling calculations performed in
the {\it Nuclear medium cooling scenario} (\citealt{bgv2004}) which differs
from the other abovementioned approaches in a consistent inclusion of medium
effects. Processes of internal heating (\citealt{tsuruta}) are not
included.
%%%%%%%%%%%%%%%%%%%%%%%%%%%%%%%%%%% end
Since the cooling history crucially depends on the assumed physical
conditions inside the star, comparison with observations may rule out
some models in favor of others.
%%%%%%%%%%%% David insertion
To study this we will vary assumptions on the nuclear pairing
gaps, the relation between crust and surface temperature as well as
presence or absence of pion condensation.
%%%%%%%%%%%% end
A customary way of testing predictions of
cooling calculations is to construct a temperature vs. age
($\mathrm T$-$\mathrm t$ for short) plot for
the largest sample of sources. Despite its wide application
and undisputed usefulness, this test has a number of limitations.
In this paper we suggest use of the $\mathrm{Log\, N}$-$\mathrm{Log\, S}$
distribution of close-by NSs as an additional probe for NS cooling models.
The idea is based on the comparison of present
observational data with NS population synthesis calculations in which cooling
curves are one of the ingredients. Our approach extends the actual
calculation of
observational properties of the population of close-by young NSs
that has already been developed by
\cite{p03,p04} (hereafter Paper I and II)
%%%%%%%%%% david insertion
by including nuclear medium effects in the cooling code and by investigating
the contribution of massive
progenitors in the Gould Belt to the local NS population
in more details.

The paper is organized as follows. In section \ref{tests} we discuss
the advantages and limitations of the two methods
in extracting information from observational data.
Section \ref{popsynth} presents our population synthesis model in more
detail, especially concerning the choice of the set of cooling curves
which will be used in our calculation.
We present our results and discuss them in section \ref{results}.
Section \ref{concl} contains our conclusions.

\section{The two tests}
\label{tests}

In this section we briefly discuss and compare the capabilities of
the conventional $\mathrm T$-$\mathrm t$ test and of the
$\mathrm{Log\, N}$-$\mathrm{Log\, S}$ test we propose here.
Our main conclusion is that the two approaches should be used together
as their advantages and disadvantages are mostly complementary to each other
especially in the case of available samples of observed objects.

\subsection{The $\mathrm T$-$\mathrm t$ test}

The $\mathrm T$-$\mathrm t$ test is the most appropriate one
to compare results of thermal evolution calculations with observations.
An important advantage of this test is that there are no additional theoretical
uncertainties except those connected with the cooling model: theoretical 
cooling
curves do not depend on unknown (or poorly known) astrophysical parameters but
only on the input physics of the star interior (for a discussion of
cooling processes and references to earlier papers, see e.g. \citealt{yak99}
and \citealt{bgv2004}). Still, there are some well known drawbacks to this 
method.

For sources associated with radiopulsars or CCOs, the star age is usually estimated from
the spindown age $P/\dot P$, or inferred from the age of the
supernova remnant.
To which extent these determinations are indeed representative
of the neutron star age is still uncertain.
The situation is even worse for {\it ROSAT\/} isolated neutron stars which are
not associated with a SNR and are with no exception radio-silent.
Current age estimates for the two brightest objects in this class
(RX J1856.5-3754 and RX~J0720.4-3125), based on dynamical considerations,
should be regarded only as guesses.

%**************************

The $\mathrm T$-$\mathrm t$ test is not very sensitive to objects with ages
$\ga 10^5$ yrs. There are two reasons for this.
First, there are only a few sources older than
this value to which the test can be applied.
Then, cooling curves sharply drop at the photon cooling
stage. Shifts between different cooling curves are comparable with data error
bars. So, it is difficult to discriminate between models that differ
mainly in this respect.

Cooling calculations provide the star temperature at the core-crust boundary
and the actual surface temperature is then obtained applying a bridging
formula (e.g. \citealt{t1979}; \citealt{y2004}).
Detailed modeling of heat transport in the highly magnetized envelope
indicates that the surface temperature may be influenced by several effects,
among which the magnetic field distribution inside the star is very important
(\citealt{gkp2004}). 
The temperature is derived from a spectral fit to the data and, as such, 
depends on the assumed emission model for the surface.
Different models (blackbody, H/heavy elements atmosphere with/without magnetic
field, solid surface) give values which can differ by a factor of a few.

Recently, it has been suggested that the Temperature-Age test should be 
refined by the introduction of a {\it brightness constraint} (BC),
which demands that cooling curves for objects within a mass interval 
anticipated as typical should not appear significantly brighter than the 
brightest of already observed objects for a given age (\citealt{g2005}, 
hereafter G05).
It has been shown in this work that some of the cooling models discussed in
BGV which would violate the BC become acceptable when the crustal 
properties are modified. 
These modifications do not affect old stars with ages $\ga 10^5$ yrs for 
which the $\mathrm{Log\, N}$-$\mathrm{Log\, S}$ test is relevant.
Due to our lack of knowledge of the mass distribution 
and atmospheric properties of cooling compact stars, we must be cautious in
applying the strong BC. 
Therefore we refer in this paper to results of the $\mathrm T$-$\mathrm t$ 
test with and without BC. 

Finally, the data sample is non-uniform: there are sources of different types,
and, more important, the sample is neither flux, nor volume limited,
and is a strongly selected one, containing objects for
which age and temperature estimates
are available\footnote{Note that
there are some young cooling NSs for which their age is not known,
in particular those not observed as active radiopulsars.
These sources
can not be included in the sample of the Temperature-Age test.}.
Clearly these sources are not representative of any real population of NSs.

\subsection{The $\mathrm{Log\, N}$-$\mathrm{Log\, S}$ test}

The $\mathrm{Log\, N}$-$\mathrm{Log\, S}$ distribution is a widely used tool
in many branches of astronomy.
For isolated NSs such an approach has been already
used by \cite{nt99} and \cite{p00b} to probe the origin of isolated NSs in the
solar proximity.
The main conclusion of these investigations is that the observed
$\mathrm{Log\, N}$-$\mathrm{Log\, S}$
cannot be easily explained assuming that the local population of NSs
originated only in the Galactic disc.
As shown in Paper I and II\nocite{p03,p04}, accounting for massive
progenitors in the Gould Belt reconciles theoretical predictions with the data.

One immediate advantage of the $\mathrm{Log\, N}$-$\mathrm{Log\, S}$ test is
that, at variance with the  $\mathrm T$-$\mathrm t$ test, no degree of
arbitrariness is introduced when observational data are analyzed: both the
fluxes and (of course) the number of sources are well measured.
In addition, this approach is a ``global'' one.
In our scenario it would not be possible to explain some particular sources
by invoking slight changes in the cooling physics.
Once the parameters of the model other than those related to the cooling
process are fixed (see section~\ref{popsynth}), a particular cooling curve
either fits the population as a whole or not. Furthermore, the
$\mathrm{Log\, N}$-$\mathrm{Log\, S}$ sample is a uniform one, i.e. objects
are flux (and probably volume) limited, and no
strong selection criteria are introduced.

For the $\mathrm{Log\, N}$-$\mathrm{Log\, S}$ test
the only necessary observational piece of information is the ROSAT
count rate. The method can
be applied to objects with unknown ages. This makes it possible
to include, for example, all the ROSAT X-ray dim NSs,
and 3EG~J1835+5918 (the Geminga twin)
in the testing sample. The $\mathrm{Log\, N}$-$\mathrm{Log\, S}$
test is mostly sensitive to NSs older than $\sim 10^5$~yrs.
Older sources dominate in number,
and in the solar proximity there are about a dozen of them in comparison
to very few with $t\la 10^5$ years.\footnote{Of course, the
$\mathrm{Log\, N}$-$\mathrm{Log\, S}$ test can be sensitive to young objects
if another uniform sample is used. For example, it is very important to make
a population synthesis of sources in SN remnants.
Then, as we mentioned above,
differences between cooling curves for large ages are tiny. However, in the
$\mathrm{Log\, N}$-$\mathrm{Log\, S}$ approach we, in some sense, integrate
these small differences along the curves, so their impact on the final
$\mathrm{Log\, N}$-$\mathrm{Log\, S}$ distribution is significant.
}

Nevertheless there are significant limitations too.
One source of uncertainty is our incomplete knowledge of some important
ingredients of the population synthesis model. These are discussed in more
detail in the next section and concern the spatial distribution of the NS
progenitors, the NS mass and velocity spectrum, and their emission properties.
However, all these issues, to some extent, can be addressed by considering
different cases believed to cover
the entire range of acceptable scenarios. In addition, there could be unknown
correlations among some of the quantities we use to parametrize our model, so
that they should not be treated as independent ones. Examples of such possible
correlations are those between the star kick and the internal structure
(because of quark deconfinement, see \citealt{bp04}),
and between the star mass and the magnetic field
(because of fallback, see \citealt{petal02, hws04}).
A more severe problem arises in connection with the low statistics of the
sample, since there are only about 20 thermally emitting NSs known to date.
This implies that the bright end of the $\mathrm{Log\, N}$-$\mathrm{Log\, S}$
relation comprises very few objects so that it is difficult to account for 
statistical fluctuations. We do not know much
either about the properties of very faint sources, i.e. the dim end of the
$\mathrm{Log\, N}$-$\mathrm{Log\, S}$ distribution.

\section{The population synthesis model}
\label{popsynth}

The main physical ingredients that enter our population synthesis model
are:

\begin{itemize}
\item the initial NS spatial distribution;
\item the kick velocity distribution;
\item the NS mass spectrum;
\item the cooling curves;
\item the surface emission;
\item the interstellar absorption.
\end{itemize}

The calculation of the NS spatial evolution as they move in the Galactic
gravitational potential follows that presented in
Papers I and II \nocite{p03, p04}. The same
treatment of the interstellar absorption is retained and the kick distribution
is that proposed by \cite{acc2002}. We do not account for atmospheric
reprocessing of thermal radiation, and assume that the emitted spectrum  is a
pure blackbody.
Although this is clearly an oversimplification, it is a reasonable
starting assumption and will serve for our, mainly illustrative, purposes.
A more detailed description of surface emission may be easily accommodated in
our model later on. For the time being, we perform our calculations for nine
different sets of cooling curves among those discussed in \cite{bgv2004},
hereafter BGV, to which we refer for all details.
This issue, together with the initial spatial distribution of NSs and their
mass spectrum, is further discussed below.

\subsection{The initial NS spatial distribution}

Following the results of previous investigations (paper~I\nocite{p03}), we
take as an established fact that the population of nearby NSs is
genetically related to the Gould Belt.
The contribution of the Belt has dominated the production of compact
remnants in the solar proximity over the past $\sim 30$~Myrs
(see \citealt{p97} for a detailed description of the Belt structure).
About two thirds of massive stars in the $\sim 600$~pc around
the Sun belong to the Belt (\citealt{torra}). Since they are bright objects,
it is possible
to  track their positions exactly for example by using HIPPARCOS data.
However, here we use a simplified distribution of the progenitor stars, as
discussed in Papers I and II\nocite{p03, p04}.
NSs are born in the Belt, for which a simple ring-shaped
structure is assumed, and in the Galactic disc. The outer belt radius is a
model parameter
and two values, 300~pc and 500~pc, were used, the second being an
upper limit. The supernova rate was taken from \cite{g2000}, and appears
in agreement with the historical rate estimated by \cite{tammann}.

\subsection{The NS mass spectrum}
\label{mass-spect}
Since cooling curves are strongly dependent on the star mass,
the mass spectrum is one of the most important ingredients and,
unfortunately, one of the lesser known.
We cannot rely on the mass measurements in binary radio pulsars
(e.g. \citealt{tho99}) because they refer to a ``twice selected''
population
(i.e. selection effects due to evolution in a binary can be
important together with possible conditions necessary for radio
pulsar formation).
Probably not all NSs
go through the active radio pulsar stage (e.g. \citealt{gv00}), and
the properties of NSs in binaries may be different to those
of isolated objects (e.g. \citealt{plp2003}).
Note that the local NS mass spectrum can be  different
to the global NS mass spectrum in the Galaxy. Even stronger deviations can
be expected between the mass spectra of local NSs and of those sources
usually used for the   $\mathrm T$-$\mathrm t$ plot.

As the population of NSs in $\sim 1$ kpc around the Sun may be slightly
different from the average galactic population, we estimated the
mass spectrum for these
objects directly (see Paper II \nocite{p04} for more details).
The basic idea is to use HIPPARCOS data on massive stars around the Sun
in conjunction with the  calculations by \cite{whw02}. Knowing the mass
distribution
of progenitors through their spectral classes, we use a fit to a plot from
\cite{whw02}
in order to obtain the NS mass from the mass of the progenitor.

\begin{figure}[t]
\vbox{\psfig{figure=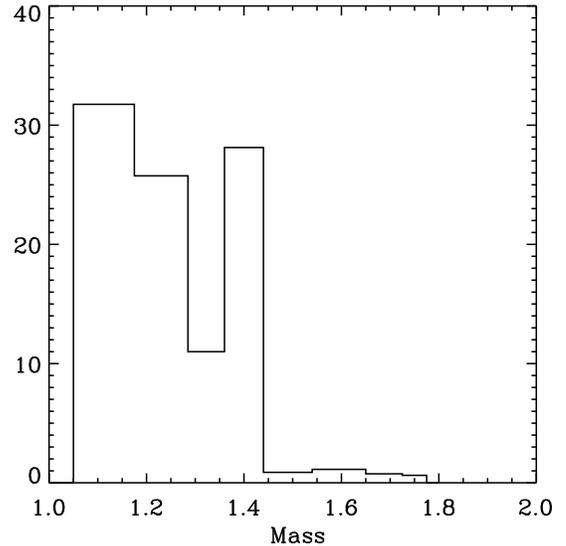,width=\hsize}}
\caption[]{The adopted mass spectrum, binned over eight intervals of
different widths.
}
\label{fig:mass}
\end{figure}

We use eight mass bins centered at
$\mathrm{M/M}_{\odot}=$ 1.1, 1.25, 1.32, 1.4, 1.48, 1.6, 1.7, 1.76.
The adopted mass spectrum is shown in Fig.~\ref{fig:mass}.
The lower limit for the NS mass is still an open question.
\cite{tww1996} suggested that there are no NSs with ${\mathrm M}\la 1.27\,
\mathrm{M_{\sun}}$, although their conclusion is not definite
(see also \citealt{whw02}).
For this reason we decide to use also a truncated mass spectrum, in
which the first bin is suppressed and all objects originally contained there
are added to the second one. Each bin corresponds to one of the calculated
cooling curves.
According to the mass spectrum, each curve has a statistical weight of
$31.75\%$, $25.75\%$, $11\%$, $28.125\%$, $0.875\%$, $1.125\%$, $0.75\%$,
and   $0.625\%$.
For the truncated one the weights of the first two bins
are replaced by 0.0 and $57.5\%$ respectively.
We note that sampling also relatively low masses is important
since low-mass NSs seem to be required to interpret data on
the $\mathrm T$-$\mathrm t$ plot (BGV).

\begin{table*}[h]
\begin{tabular}{|c|c|c|c|c|c|c|c|c}
\hline
Model&Reference &$\pi$ Cond& Gaps & Crust&\multicolumn{2}{c|}{T -- t} & 
Log N -- Log S\\
& && & &wo BC&w BC & \\
%& &condensate&&& (G)\\
\hline
\hline
I   & BGV, Fig. 21 & Yes & A & C & Yes & Yes & Yes\\
II  & BGV, Fig. 13 & No  & B & D & Yes & No & No\\
III & BGV, Fig. 15 & Yes & B & C & Yes & No & No \\
IV  & BGV, Fig. 12 & No  & B & C & Yes & No & No\\
V   & BGV, Fig. 16 & Yes & B & D & Yes & No & No\\
VI  & BGV, Fig. 14 & No  & B & E & Yes & Yes & No\\
VII & BGV, Fig. 18 & Yes & B'& C & Yes & No & No\\
VIII& BGV, Fig. 19 & Yes & B''&C & Yes& Yes & Yes\\
IX  & BGV, Fig. 20 & No  & A  & C & Yes & Yes & Yes \\
%\hline
X   & G05, Fig. 2 & Yes & B & E &  Yes & Yes & No\\
XI  & G05, Fig. 2 & Yes & B'& E &  Yes & Yes & No\\
\hline
%\hline
\end{tabular}
\vspace{5mm}

\caption{Properties of the selected cooling curves: A - gaps from
\cite{tt2004}, $^3{\mathrm P}_2$ neutron gap suppressed by 0.1; B
- gaps from \cite{y2004}, $^3{\mathrm P}_2$ neutron gap suppressed
by 0.1; B' - same as for B and $^1{\mathrm P}_0$ proton gap
suppressed by 0.5; B'' - same as for B,  $^1{\mathrm P}_0$ proton
gap suppressed by 0.2 and $^1{\mathrm P}_0$ neutron gap suppressed
by 0.5; C - $T_{\rm s} - T_{\rm in}$ relation fit from BGV; D -
$T_{\rm s} - T_{\rm in}$ relation by \cite{t1979}; E - $T_{\rm s}
- T_{\rm in}$ relation from \cite{y2004} and $\eta = 4\times
10^{-6}$. {The last three entries indicate whether the
model complies or not with the tests: temperature-age without
(wo) or with (w) the additional brightness constraint and
$\mathrm{Log\, N}$-$\mathrm{Log\, S}$ 
(see text and also BGV and G05 for more details)}.
\label{tab:models}}
\end{table*}

\subsection{Cooling curves}
\label{cooling}

In their recent paper \cite{bgv2004} presented sixteen sets of cooling curves.
Each set contains models for several values of the star mass while different
sets refer to different assumptions on heat transport in the crust and on the
physical processes in the NS interior.
Five of these models are unable to reproduce the observed
temperature-age plot and will not be considered further.
From the remaining eleven sets, all of which give results not in contradiction
with observations
(see, however, the discussion in sec. \ref{results}), we select nine
representative models for our population synthesis calculations
(models I - IX in Table \ref{tab:models}).
We add two models (X and XI) from the recent analysis in G05, 
which correspond to models III and VII, respectively, when calculated 
with different crustal properties.
All of them have superfluid nuclear matter and medium modifications of the
neutrino processes.
They differ in the assumptions about the superfluid gaps, the presence/absence
of a pion condensate and the properties of the neutron star crust.
The latter governs the relationship between the temperature
of outermost core layer ($\mathrm{T_{in}}$) to that of the star surface
($\mathrm{T_{s}}$).
The main characteristics of the selected models are summarized in
Table \ref{tab:models}. All of the eleven models satisfy the temperature - age
test according to BGV, whereas only six of them fulfill the additional
brightness constraint  introduced in G05\nocite{g2005}. {The last
column anticipates the results discussed in the following section
and shows if the model complies with the 
$\mathrm{Log\, N}$-$\mathrm{Log\, S}$ test}.

\section{Results and discussion}
\label{results}

% MODEL I
\begin{figure*}
\hbox{\psfig{figure=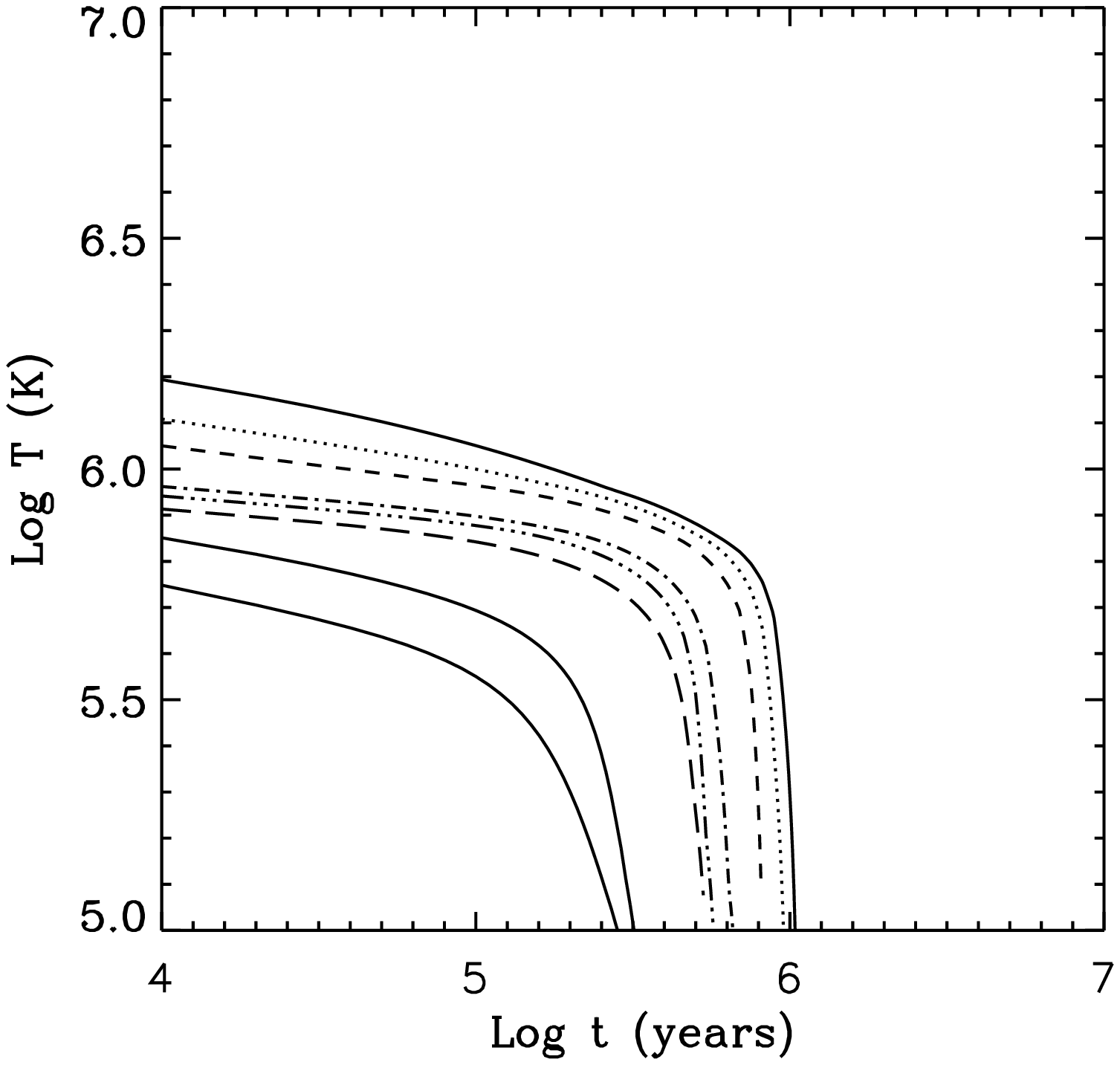,width=9cm}
\psfig{figure=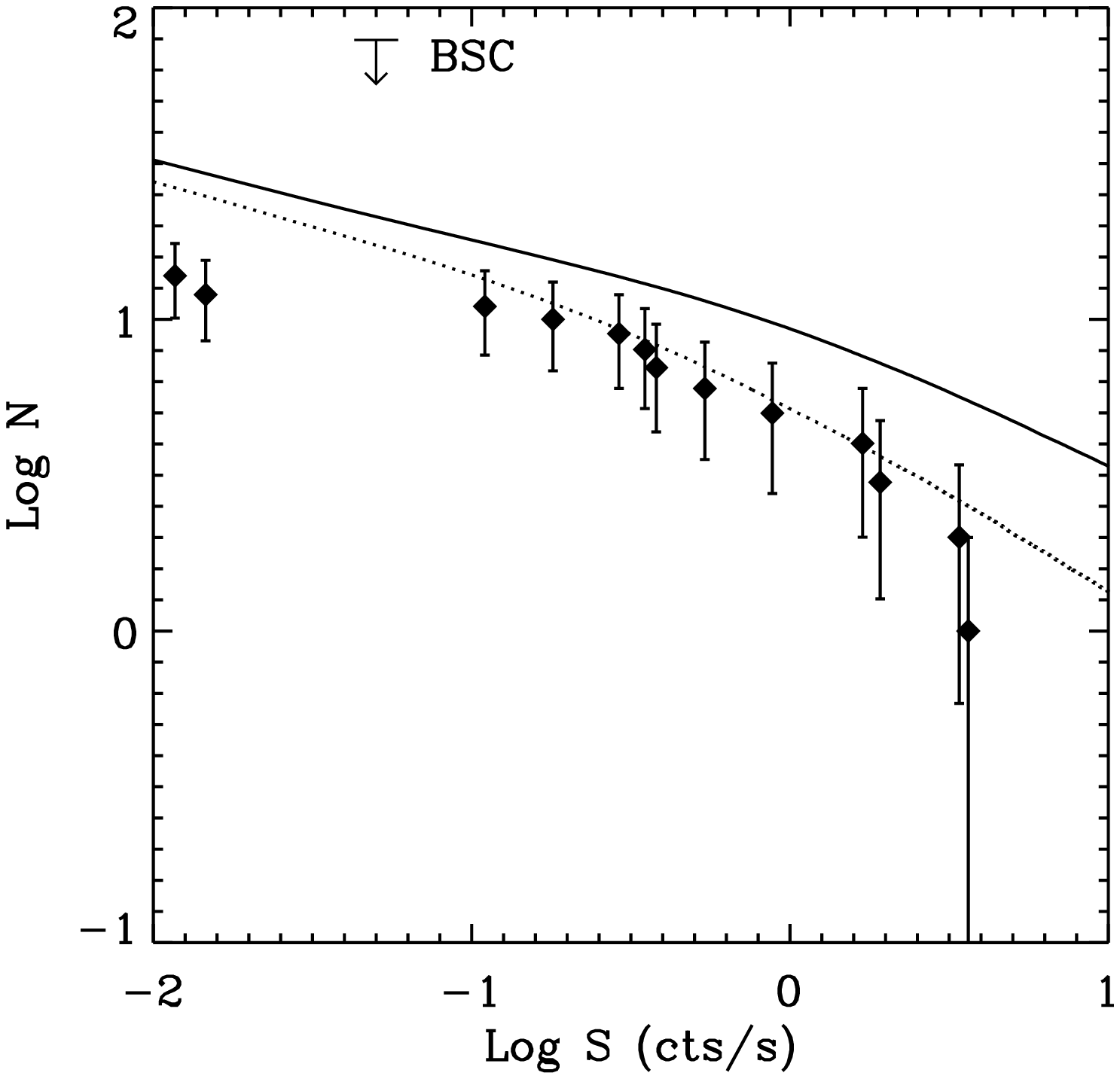,width=9cm}}
\caption[]{ Model I. Left: cooling curves for (from top to bottom)
1.1, 1.25, 1.32, 1.4, 1.48, 1.6, 1.7, 1.76 $\mathrm{M}_\odot$. Right: the
corresponding $\mathrm{Log\, N}$-$\mathrm{Log\, S}$ distribution for
$\mathrm{R_{belt}}=300$~pc and non-truncated mass spectrum (full line)
and 500~pc and truncated mass spectrum (dotted line). See text
for details.
\label{fig:m1}}
\end{figure*}
%
% MODEL III
\begin{figure*}
\hbox{\psfig{figure=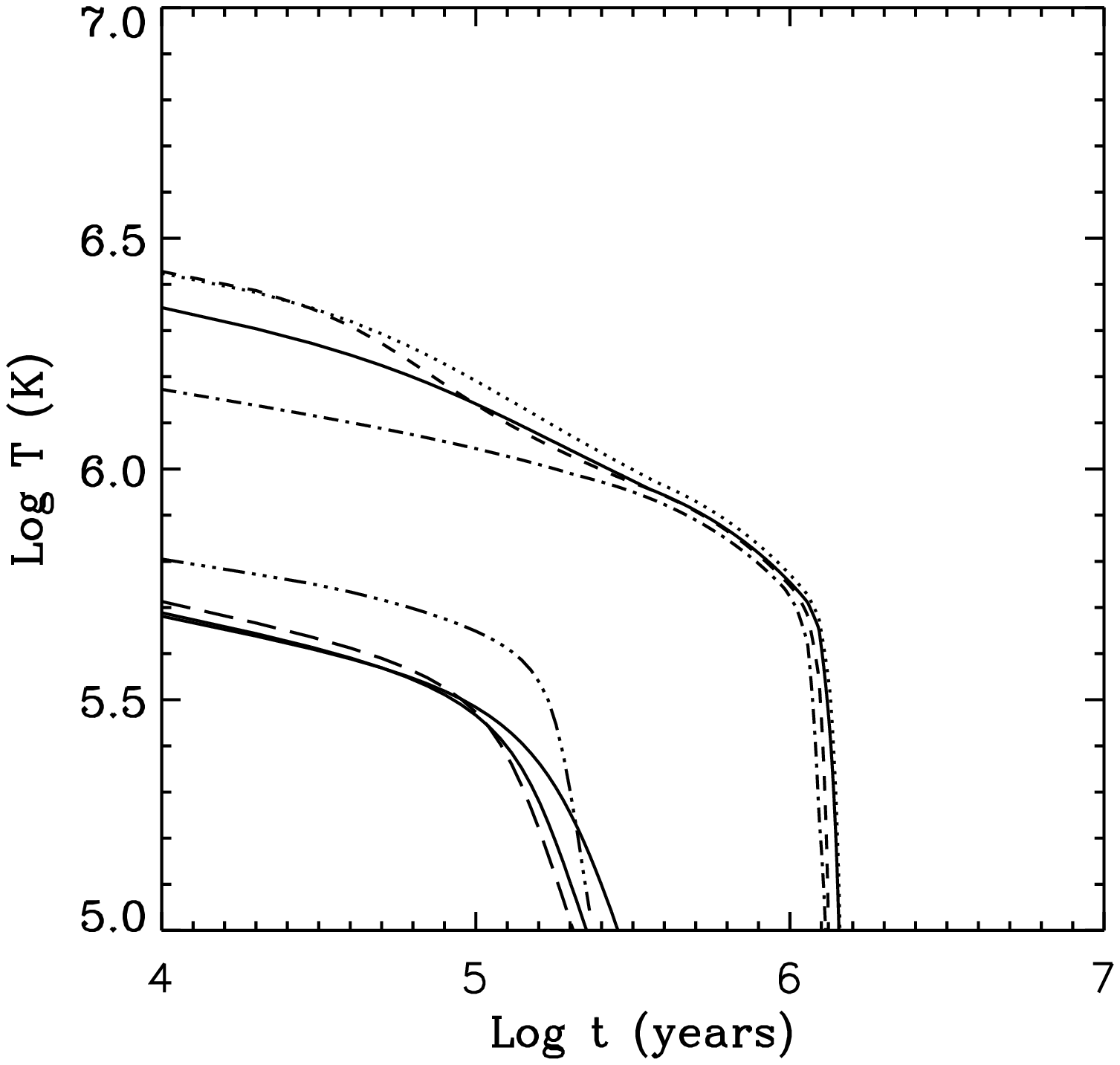,width=9cm}
\psfig{figure=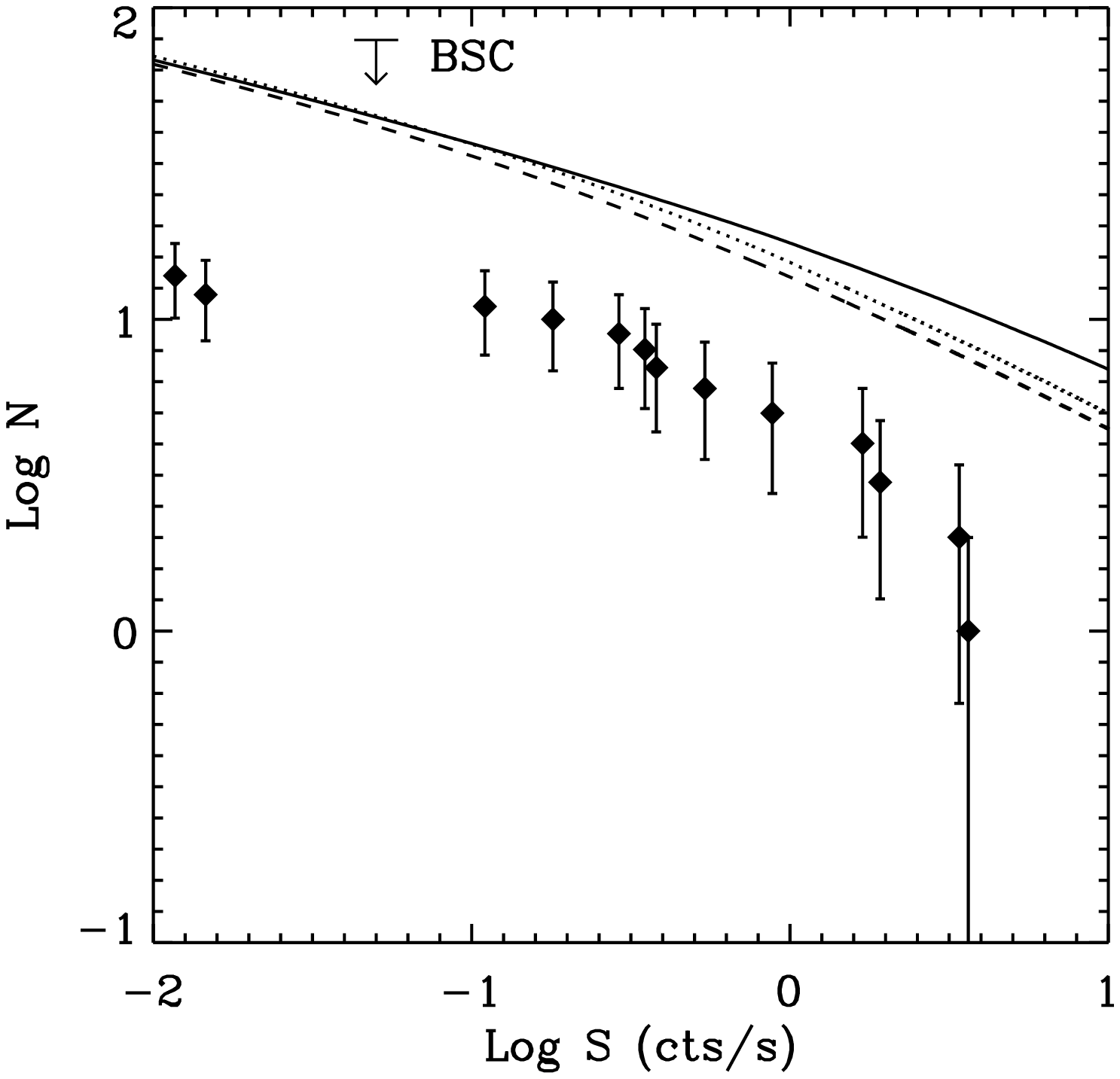,width=9cm}}
\caption[]{Same as in Fig. \ref{fig:m1} for Model III. The dashed line
refers to a calculation in which the full (non-truncated)
mass spectrum was used and $\mathrm{R_{belt}}$ was assumed to be $500$~pc.
Model X should produce nearly the same $\mathrm{Log\, N}$-$\mathrm{Log\, S}$
distribution as it differs only in the type of the crust.
% and should be compared with the dotted one.
\label{fig:m3}}
\end{figure*}
%
% MODEL VI
\begin{figure*}
\hbox{\psfig{figure=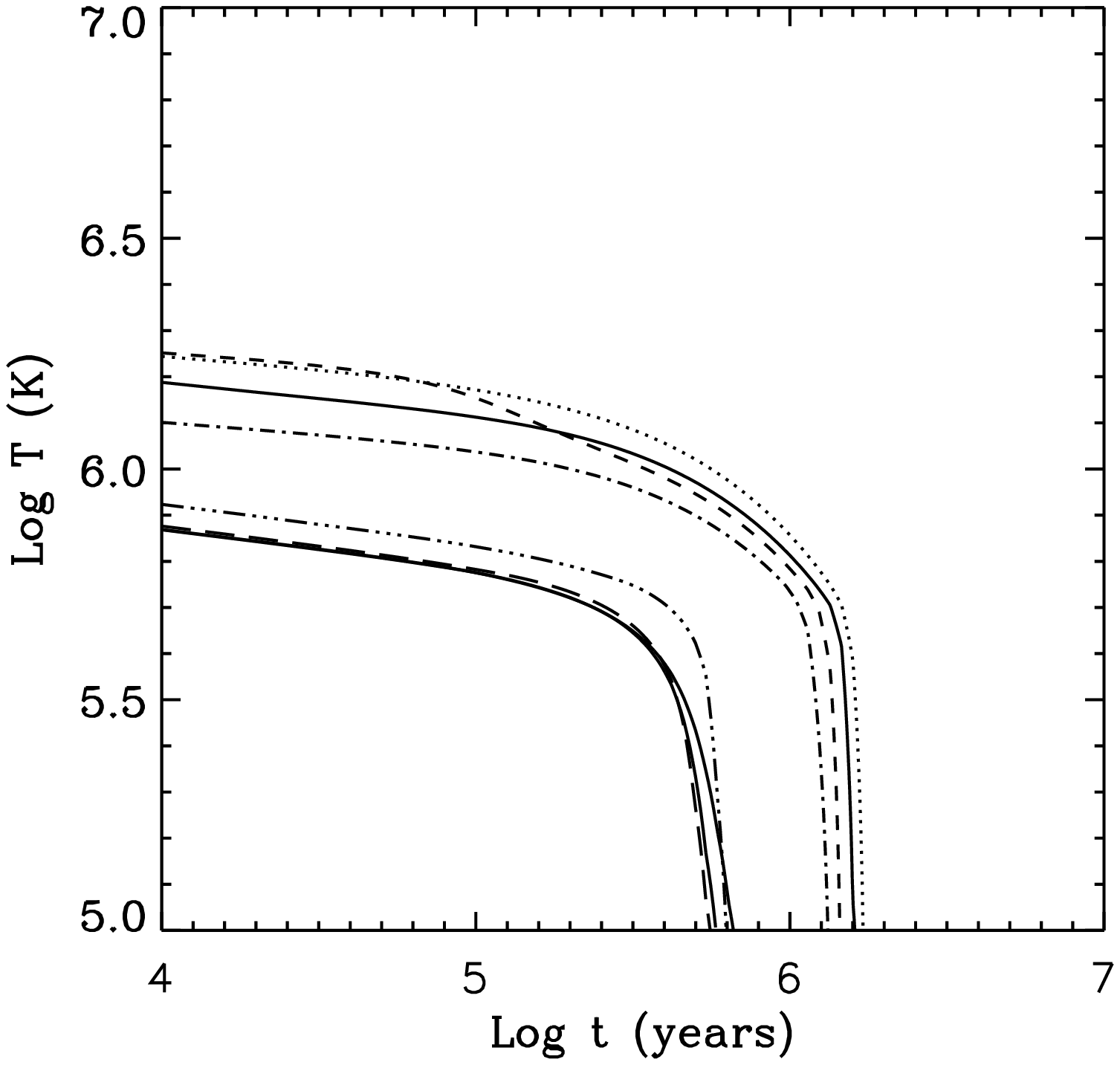,width=9cm}
\psfig{figure=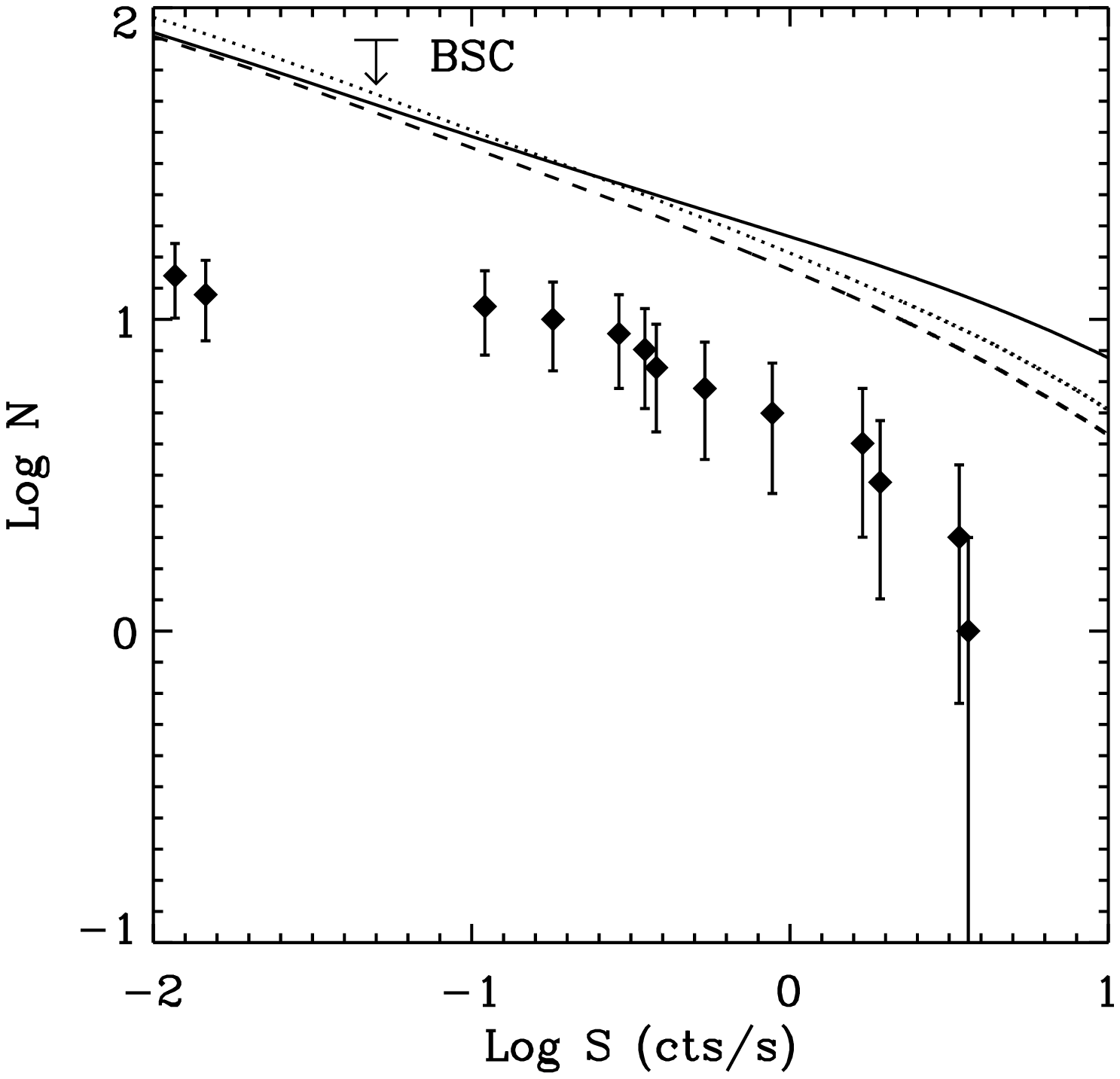,width=9cm}}
\caption[]{Same as in Fig. \ref{fig:m3} for Model VI.
{Model IV gives quite similar results since the change of the crust model
from C to E does not affect the log N - log S distribution}.
\label{fig:m6}}
\end{figure*}
%
% MODEL VII
\begin{figure*}
\hbox{\psfig{figure=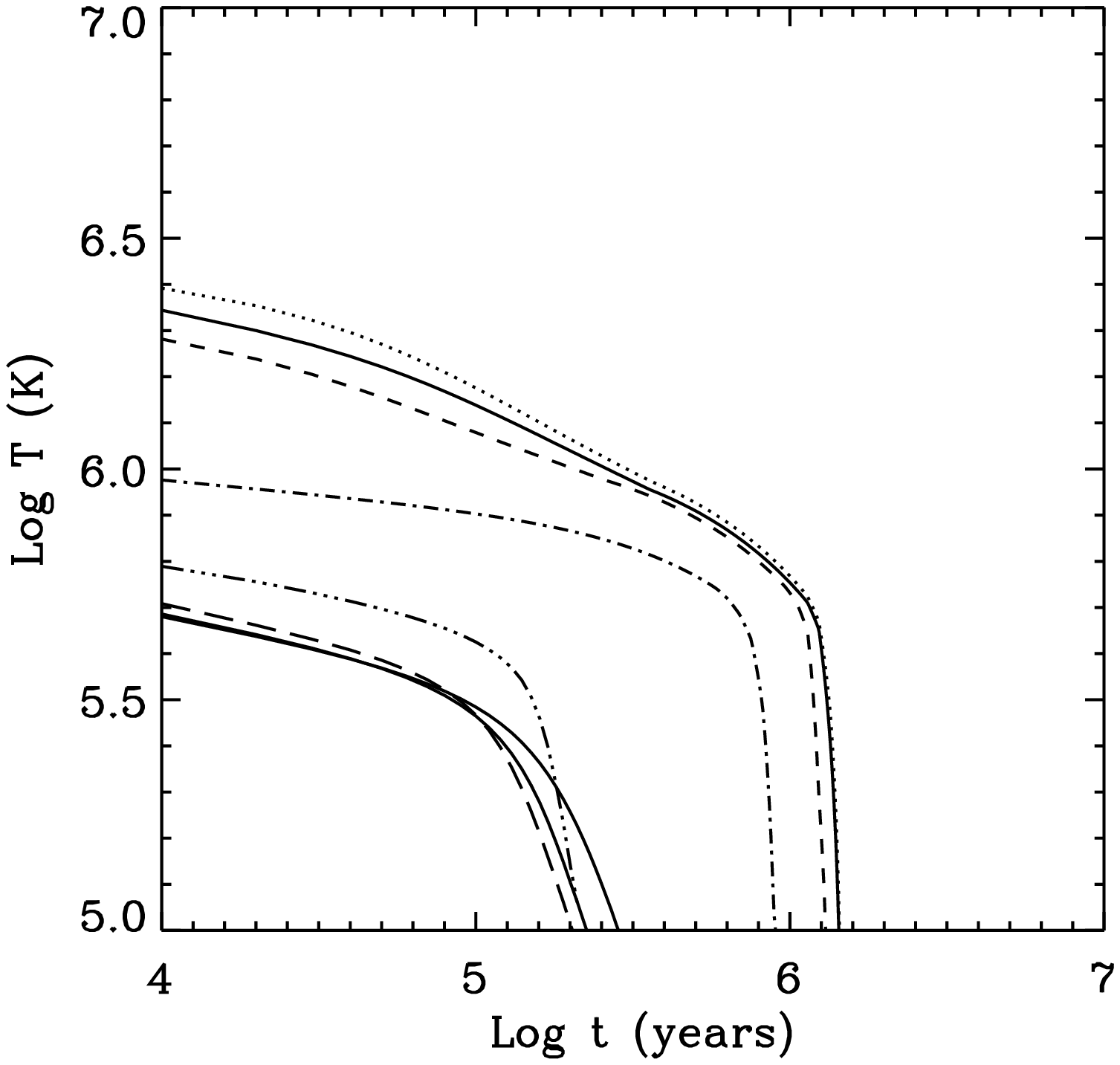,width=9cm}
\psfig{figure=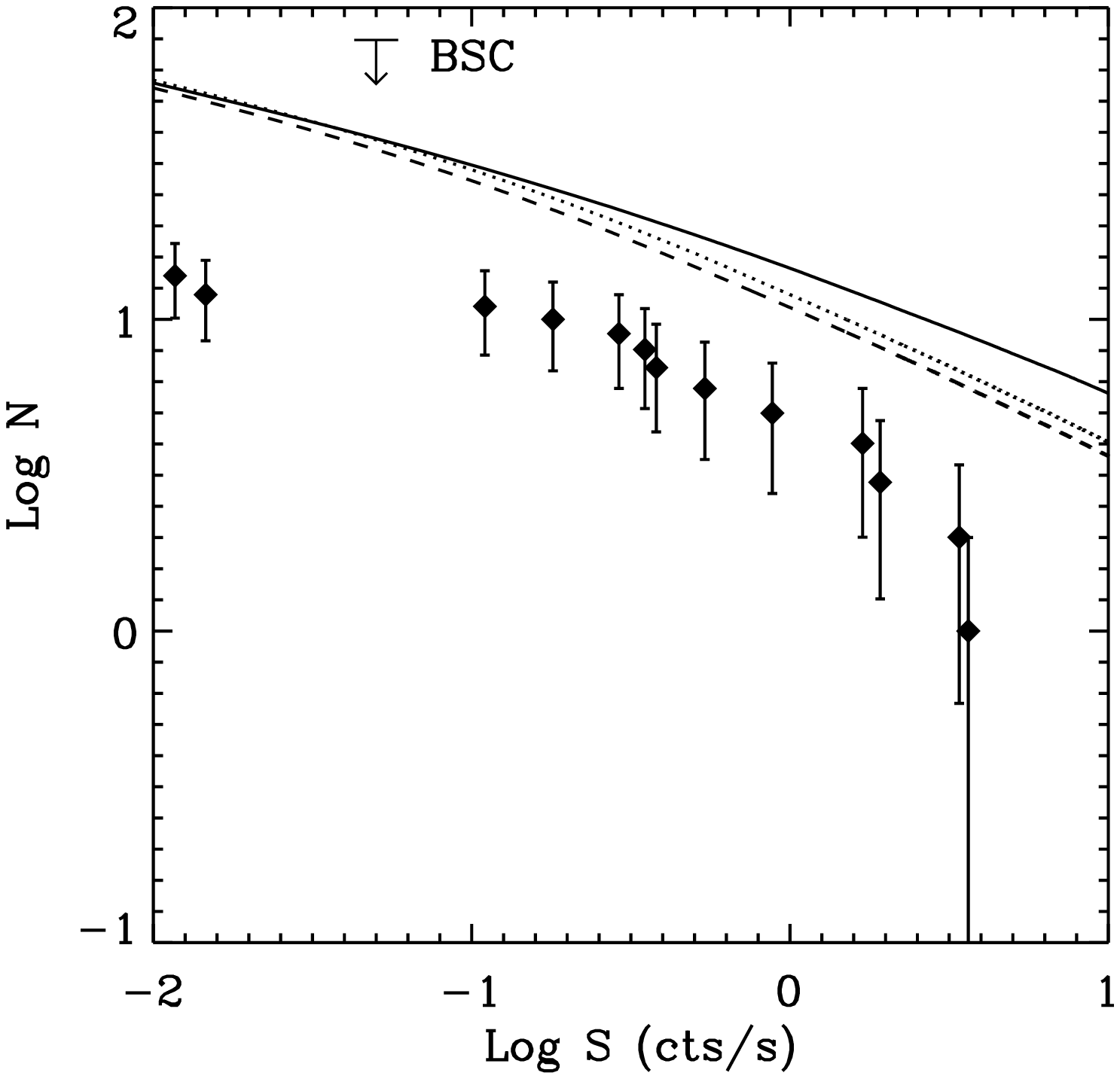,width=9cm}}
\caption[]{Same as in Fig. \ref{fig:m3} for Model VII.
Model XI should produce nearly the same $\mathrm{Log\, N}$-$\mathrm{Log\, S}$
distribution as it differs only in the type of the crust.
\label{fig:m7}}
\end{figure*}
%
% MODEL VIII
\begin{figure*}
\hbox{\psfig{figure=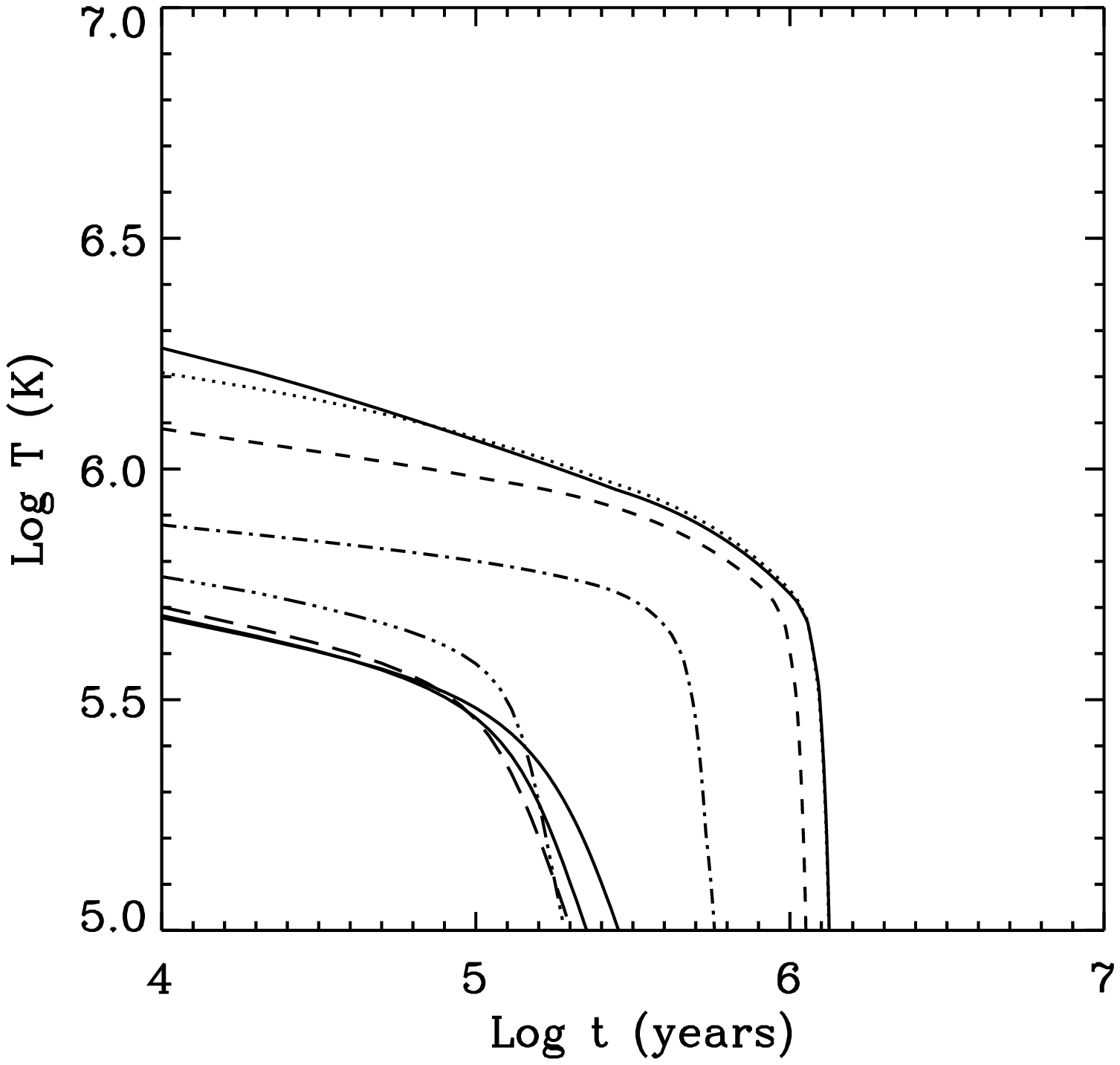,width=9cm}
\psfig{figure=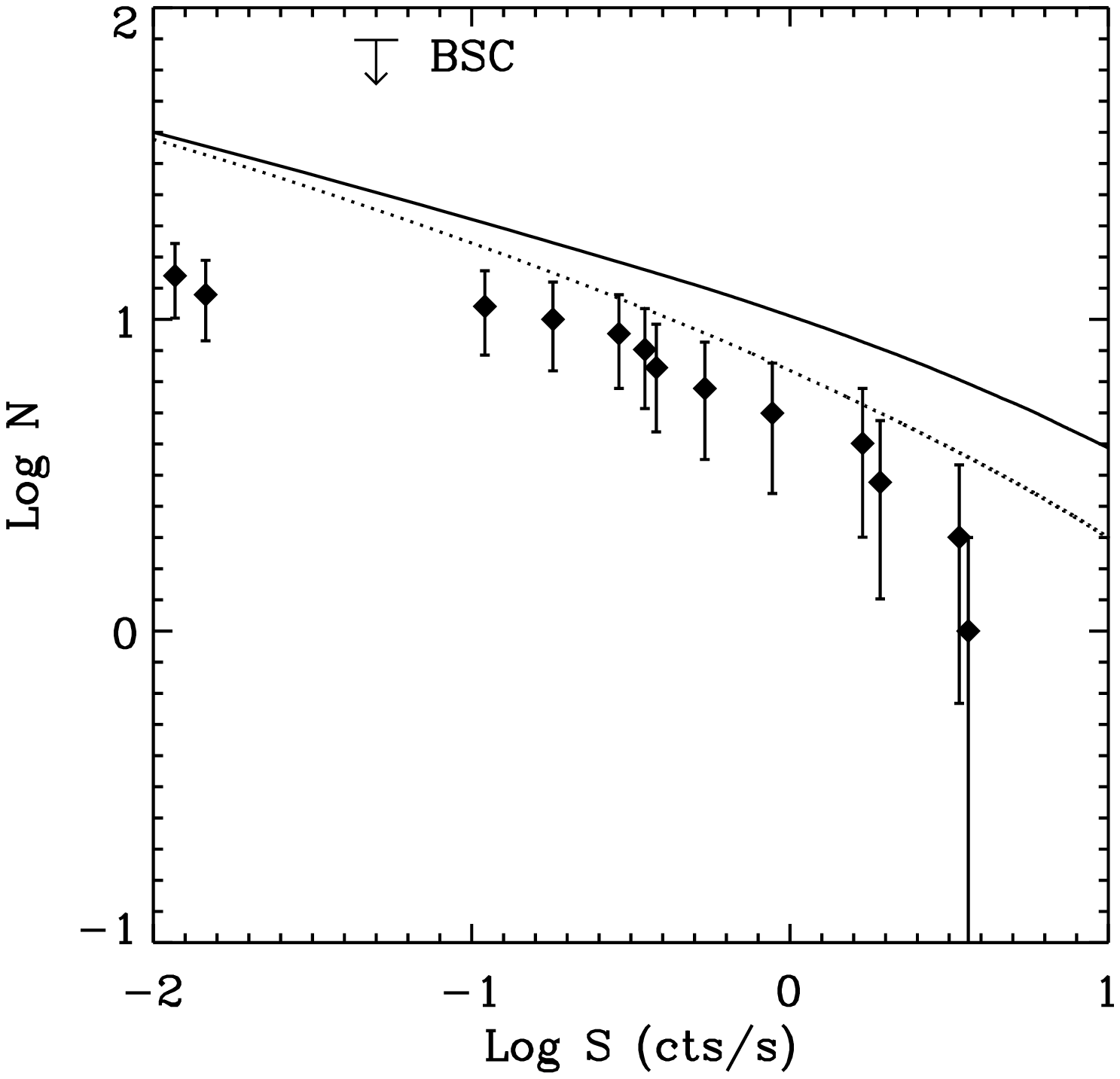,width=9cm}}
\caption[]{Same as in Fig. \ref{fig:m1} for Model VIII.
\label{fig:m8}}
\end{figure*}
%
% MODEL IX
\begin{figure*}
\hbox{\psfig{figure=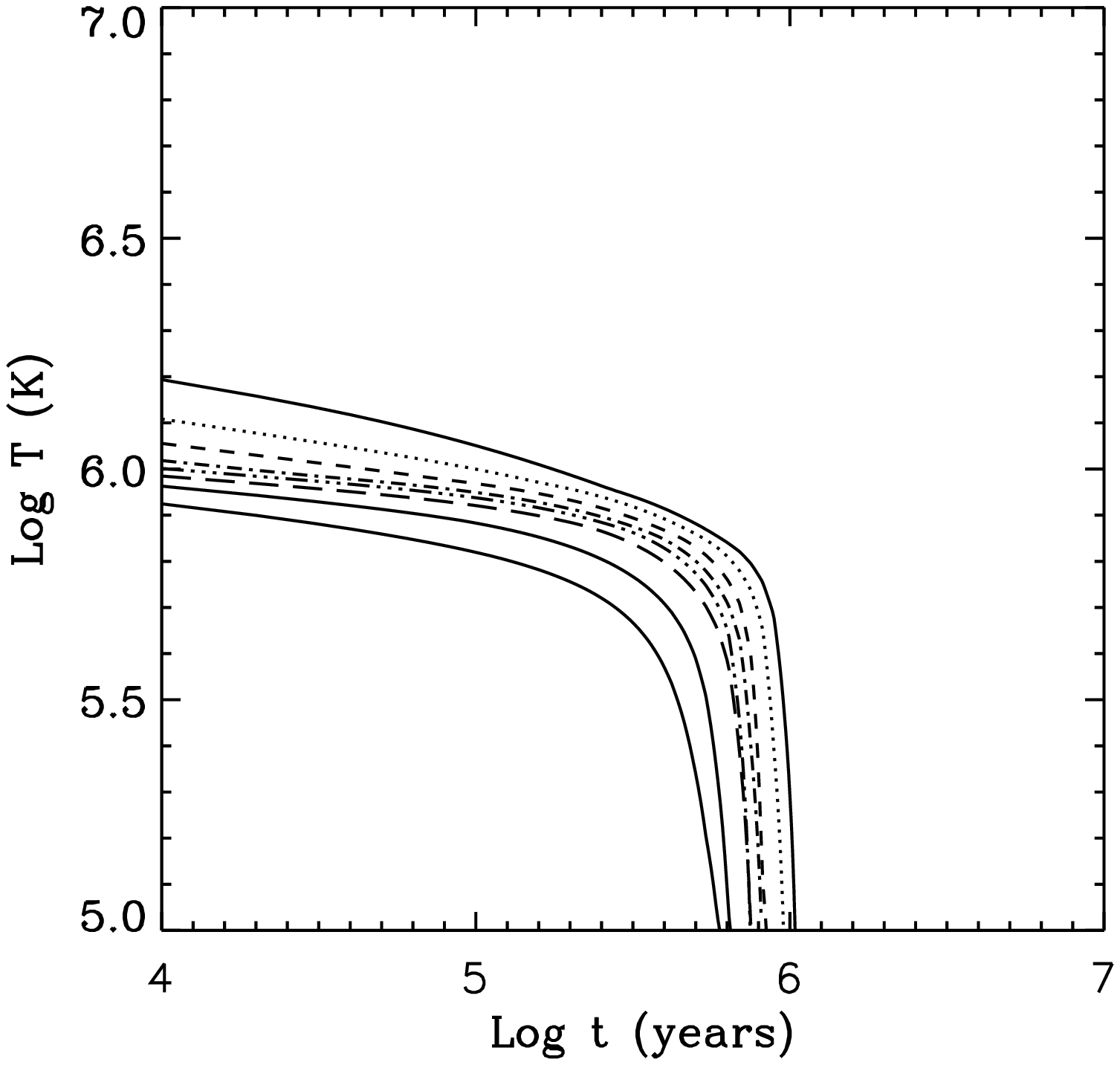,width=9cm}
\psfig{figure=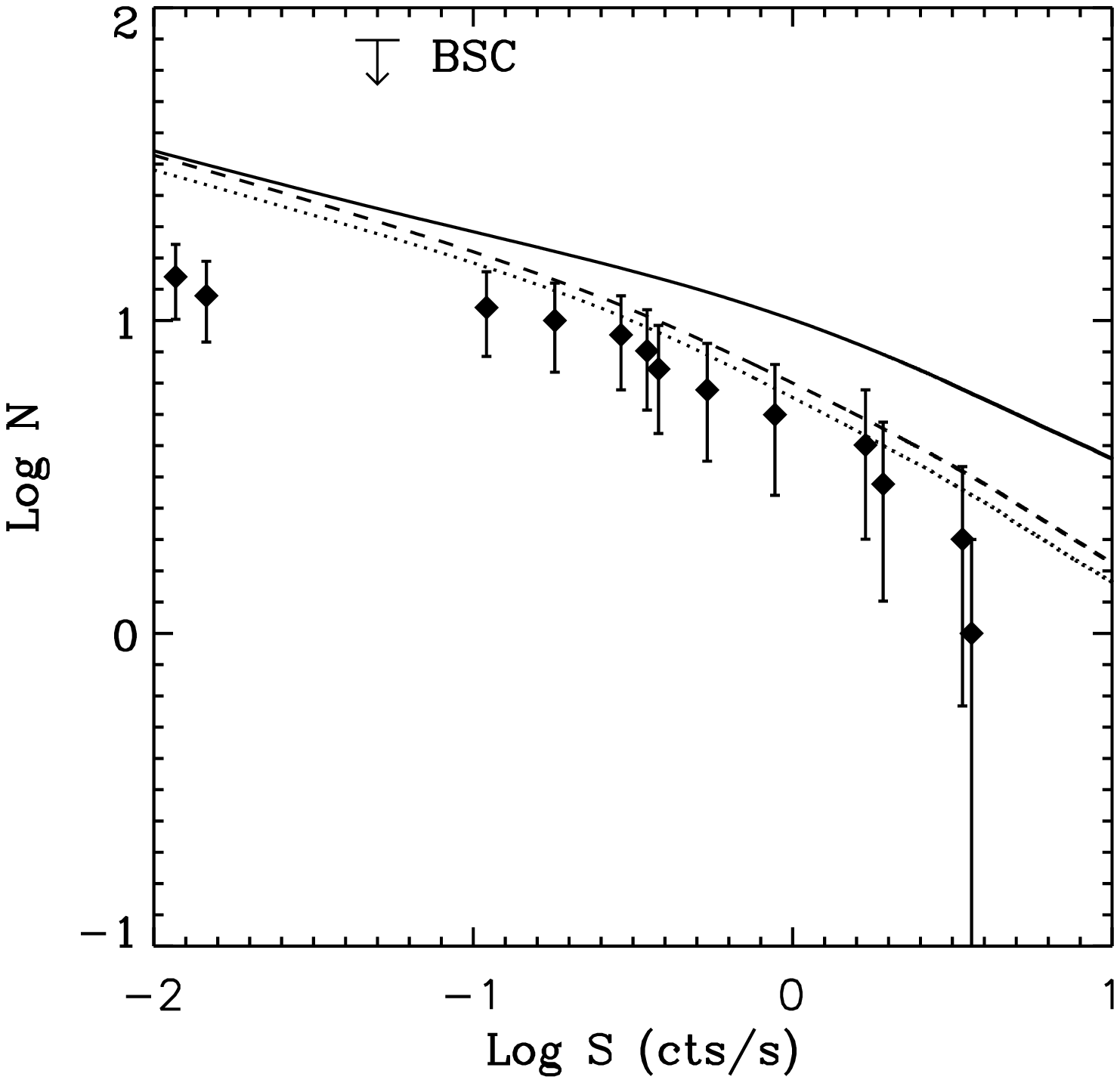,width=9cm}}
\caption[]{Same as in Fig. \ref{fig:m3} for Model IX.
\label{fig:m9}}
\end{figure*}

In each run we calculate 5000 individual tracks for the spatial evolution of a
single star with a time step of $10^4$~yrs.
Each track is applied to all eight (or seven for the truncated spectrum)
masses, and the thermal evolution is followed by the correspondig cooling
curve. Results are then collected according to the statistical
weight of each mass bin.

Results are summarized in Table \ref{tab:models} and 
Figs.~\ref{fig:m1}-\ref{fig:m9}, which refer to a selected
sub-sample  of the cooling curve sets listed in 
that Table.
In the left panels of Figs.~\ref{fig:m1}-\ref{fig:m9}, the corresponding 
cooling curves for the various masses are shown.
Results are plotted for ages $> 10^4$ yrs and temperatures above
$10^5$~K. The right panels illustrate the $\mathrm{Log\, N}$-$\mathrm{Log\, S}$
distributions computed for the same sets of cooling curves.
All models have been calculated using both the full mass spectrum and the
truncated one, although results for the latter are not shown in all cases
(see the following discussion).
Two values of the outer radius of the Gould Belt have been used to test the
dependence of our calculation
on the assumed geometry, $R_\mathrm{belt}=300$ pc and 500 pc.

Theoretical distributions are superimposed on the observed
$\mathrm{Log\, N}$-$\mathrm{Log\, S}$
for isolated NSs. The data points are derived from the sample
of thirteen sources listed in Paper~I\nocite{p03}.
Error bars correspond to Poisson statistics and are plotted to
illustrate the statistical significance of
the points. We note, however, that there can be more
unidentified sources, especially at fluxes below 0.1 cts~s$^{-1}$ (see
\citealt{rut2003} for a recent discussion). In this respect, the fact that the
two points at the lowest fluxes lie below the general trend of the observed
$\mathrm{Log\, N}$-$\mathrm{Log\, S}$ is not surprising.
In addition, the observational
upper limit derived by \cite{rut2003} on the number of fainter sources
is also shown (marked as BSC in the figures). 
Bounds on the total number of sources with flux $> 0.2$~cts~s$^{-1}$ have
been also presented
by \cite{s99}, on the basis of the {\it ROSAT\/} Bright Sources
Catalogue (BSC).

The comparison of the predicted and observed
$\mathrm{Log\, N}$-$\mathrm{Log\, S}$
distributions in Figs.~\ref{fig:m1}-\ref{fig:m9} indicates that at most three
cooling models (model I and possibly models VIII and IX)
are in agreement with the data.
All the others substantially overpredict the observed number of sources
at all fluxes, even though they comply with the  $\mathrm T$-$\mathrm t$ test.
This latter statement deserves further comment.
BGV did not reject models II--VII  on the basis of
the fact that the cooling curves cover, for the assumed mass range, the entire
region in the $\mathrm T$-$\mathrm t$ plane where the observed sources lie.
This approach is sound, and it represents the only possible option
to discriminate among different cooling scenarios at the zero level, i.e
without introducing additional information.
Clearly, if some assumptions on the NS mass distribution are
made, the $\mathrm T$-$\mathrm t$ test can be used to exclude some further
models.
If the same mass distribution discussed in  Sec. \ref{mass-spect} is applied to
BGV sets of cooling curves, one is immediately led to discard models II--V
and VII because they predict quite high temperatures for low mass NSs
($\mathrm M\la 1.3\mathrm M_\odot$)
which, according to our mass spectrum, are very abundant.
Although such hot objects would be
detectable even to large distances they are not actually observed
(this issue is further discussed in G05, see also below).
However, care must be taken in using such an argument.
Our mass spectrum is meant to be representative
of the local population of
isolated NSs, and its application to the very biased and limited
sample of objects which can be placed on the
$\mathrm T$-$\mathrm t$ plots is uncertain.
Nevertheless, one can reverse the argument by saying that the two tests,
{\em when provided with the same amount of information},
yield results that are broadly consistent, as they should.
Still, we note that a strict interpretation of the $\mathrm{Log\, N}$-
$\mathrm{Log\, S}$ test
results in the exclusion of two further models (VI and VIII) and that even
in the most optimistic case model VI is rejected.

The $\mathrm{Log\, N}$-$\mathrm{Log\, S}$ test is not equally sensitive
to changes in the three main groups of parameters in Table \ref{tab:models} 
(presence or absence of the pion condensate, gaps, and type of the crust). 
Mostly the test reacts to changes in the gap parameters. 
Conversely, changes in the type of the crust are not very important as here 
we consider a sample of relatively old NSs (nevertheless uniform samples 
of younger NSs can also be studied).
Variations in the crust properties are discussed in detail in G05, where the 
{ brightness constraint\/} was introduced.
G05\nocite{g2005} shows that models II--V and VII can be ruled out when the 
absence of very bright young NSs is considered as a constraint. 
The fact that the $\mathrm{Log\, N}$-$\mathrm{Log\, S}$ test leads to the 
same conclusion is however a completely independent result, since objects 
from different age ranges are used in the two approaches.

As mentioned earlier, the $\mathrm{Log\, N}$-$\mathrm{Log\, S}$ test applied 
to close-by NSs is not sensitive to changes in the crustal properties. 
This implies that if a cooling model with
a given  crust is rejected then its variants with other types of crust
(at least from the set considered here) can be ruled out, too. 
For example, additional models X and XI (see Table \ref{tab:models}) do not 
satisfy the test because their {\it twins} (models III and VII, respectively) 
do not (see Figs. \ref{fig:m3} and \ref{fig:m7}). 
In G05\nocite{g2005} it has been shown that it is possible to fulfil the 
BC by changing the crust. 
For example, model III with crust E can fit the data. 
However, on the basis of the $\mathrm{Log\, N}$-$\mathrm{Log\, S}$ test this 
model is ruled out.
This is because the latter test mainly ``feels'' the changes in gaps. And
if they are ``wrong'' then changes in the crust properties are of no avail.

Jumping to the conclusion that the present analysis can provide direct 
information on the physical state of star interior would be premature.
Our adopted scenario contains a number of uncertainties, as discussed in
sections~\ref{tests} and \ref{popsynth}.
Nevertheless, we believe that the case presented here convincingly shows
that by combining theoretical cooling curves with populations synthesis
calculations one has the potential to discriminate between competing cooling
scenarios.
The main outcome of this investigation is not that models
I, and possibly VIII and IX, fit the data while others do not.
This may be the result of our starting assumptions.
What matters is the fact that, within the same set of assumptions,
different cooling models
produce different results when compared with observations.

Current limitations of this approach are due to our present incomplete
knowledge of some key issues, chiefly the NS mass spectrum, their surface
emission properties and initial spatial distribution.
We attempted to account for some of these uncertainties
in our model by considering different configurations which should bracket the
true behavior.  We note that in all the six cases for which no agreement
has been found between theoretical and observed distributions,
this is largely independent of the assumptions we introduced. Using different
values of $R_{\mathrm{belt}}$ or taking variants of the mass spectrum does not help
in reconciling predictions with the data. However, we caution that other
effects, like those introduced by the proper inclusion of an atmospheric
model, may be important. This, and other issues, will be the subject of 
future work.

In all the cases we examined (see again Figs.~\ref{fig:m1}-\ref{fig:m9}),
with the possible exception of model VIII, the capability of our test to
discriminate between different cooling scenarios seems to be quite robust.
The three models (I, VIII and IX) which can reproduce the observed
$\mathrm{Log\, N}$-$\mathrm{Log\, S}$ for our choice of parameters have also
been considered by BGV as the most realistic ones.
Among models I--IX which according to BGV are not in contradiction
with the Temperature--age test,
the three
mentioned above are the theoretically most appealing ones since either
the superfluid gaps were calculated with
the same nucleon-nucleon interaction that formed the basis for the equation
of state and thus the structure of the neutron star configurations
(models I and IX) or the gaps were modelled such as to mimic these results
(model VIII).

\section{Conclusions}\label{concl}

In this paper we suggest adopting the $\mathrm{Log\, N}$-$\mathrm{Log\, S}$ 
test as an useful addition to the standard $\mathrm T$-$\mathrm t$ test in 
probing neutron star cooling models. 
To illustrate the capabilities of the proposed approach, 
we applied it to nine sets of cooling curves from \cite{bgv2004}.
Out of sixteen sets described in that paper
and two additional ones taken from \cite{g2005}, these eleven produce
results that are not in immediate contradiction with the
$\mathrm T$-$\mathrm t$ test.
The application of the $\mathrm{Log\, N}$-$\mathrm{Log\, S}$ test rules out
at least eight out of eleven investigated cooling models,
resulting in just three models able to pass both tests.
Requiring that the tested cooling models should fulfill in addition
the more stringent bightness constraint of G05, there are still six
models left out of which only 50\% pass the
$\mathrm{Log\, N}$-$\mathrm{Log\, S}$ test. 

One of the most challenging questions for the application of the test
suggested in this paper is to use it next for a possible discrimination
between purely hadronic compact star cooling scenarios and hybrid ones for
stars having a color superconducting quark core \cite{gbv2005} which have
already successfully passed the T-t test.
Our conclusion is that the $\mathrm{Log\, N}$-$\mathrm{Log\, S}$ may
therefore become a powerful strategy in uncovering the properties
of dense nuclear matter under the extreme conditions in neutron star
interiors.

\begin{acknowledgements}
We thank D.~Yakovlev and other members of the St-Petersburg group
for many helpful discussions. We are indebted to D.~Voskresensky
and M.~Prokhorov for their constant interest and support.
Comments by the referee were very useful and helped to improve the paper.
S.P. acknowleges a postdoctoral fellowship from the University of Padova where
most of this work was carried out and a fellowship from
the ``Dynasty'' Foundation. The work of S.P. was partly supported through
RFBR grant 03-02-16068.
H.G. is grateful to the Department of
Physics, University of Padova, for hospitality and acknowledges financial
support from the Virtual Institute ``Dense hadronic matter and QCD phase
transition'' of the Helmholtz Association (grant No. VH-VI-049) and from 
Deutsche Forschungsgemeinschaft (grant No. 436 ARM 17/4/05).
The work was partially supported by the Italian
Ministry for Education, University and Research under grant PRIN-2002-027245.

\end{acknowledgements}

\end{document}